\documentclass[final]{egpubl}
\makeatletter 
\providecommand{\p@shortauthor}{} 
\providecommand{\p@copyrightTextShort}{} 
\providecommand{\p@copyrightTextShortEven}{} 
\providecommand{\p@copyrightTextTitPag}{} 
\makeatother 

\usepackage{sca2026}
\SpecialIssuePaper

\usepackage[T1]{fontenc}
\usepackage{dfadobe}  

\usepackage[
    backend=biber,
    bibstyle=EG,
    citestyle=alphabetic,
    backref=true
]{biblatex}
\usepackage[pdftex]{graphicx} \pdfcompresslevel=9

\usepackage{url}
\usepackage{hyperref}
\usepackage{egweblnk} 

\usepackage{amssymb}
\usepackage{mathtools}
\usepackage{booktabs}
\usepackage{tabularx}
\usepackage{siunitx}
\usepackage{multirow}
\usepackage{amsmath}
\usepackage{pifont}
\usepackage{makecell}

\usepackage{xcolor}

\newcommand{\figref}[1]{Figure~\ref{#1}}

\newcommand{\secref}[1]{Section~\ref{#1}}
\newcommand{\tabref}[1]{Table~\ref{#1}}

\newcommand{\cmark}{\ding{51}}%
\newcommand{\xmark}{\ding{55}}%

\newcommand{\rev}[1]{{\color{black} #1}}

\title[Two2Four: Generative Quadruped Puppeteering from Human Motion]%
      {Two2Four: Generative Quadruped Puppeteering \\ from Human Motion}

\author[Zargarbashi et. al.]
{\parbox{\textwidth}{\centering 
Fatemeh Zargarbashi$^{1,2,*}$, 
Zehong Qiu$^{1,*}$,
Dhruv Agrawal$^{1,2}$, 
Stelian Coros$^{2}$, 
Robert W. Sumner$^{1,2}$, \\
Martin Guay$^{1}$, 
Jakob Buhmann$^{1}$
}
\\
\centering 
 $^1$DisneyResearch|Studios, Switzerland
 \quad
 $^2$ETH Z\"urich, Switzerland
 \quad 
 $^*$Equal contribution
}

\begin{document}

\teaser{
    \centering
    \includegraphics[width=\linewidth]{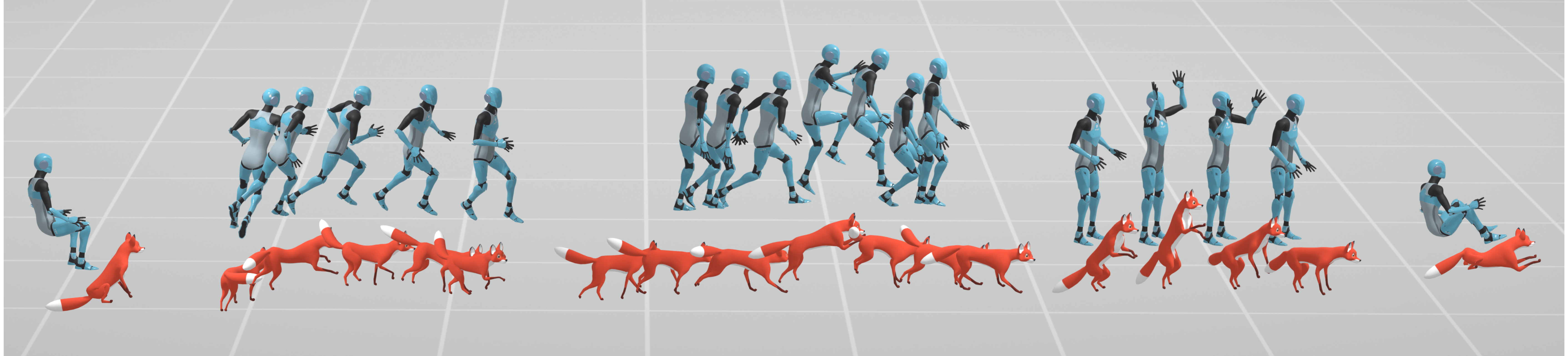}
    \caption{Given ordinary human motion, Two2Four generates realistic and controllable quadruped motion via a two-stage diffusion model, supporting diverse behaviors and fine-grained control through structured conditioning and inpainting.}
\label{fig:teaser}
}

\maketitle
\begin{abstract}
   Realistic animal motion for virtual production is typically obtained either through motion capture of highly trained performers who accurately mimic animal behavior, or by retargeting ordinary human motion using complex control setups.
   Both approaches are challenging and often fail to fully reproduce the nuances of natural animal motion, motivating data-driven alternatives.
   We present an automatic human-to-quadruped puppeteering framework that produces plausible and controllable quadruped motions from ordinary human motion data. 
   Our approach employs a two-stage generative diffusion model trained purely on quadruped motion data. 
   By introducing a structured conditioning and inpainting strategy, our method supports a wide range of actions, including walking, running, jumping, sitting, and lying.
   Furthermore, we enable fine-grained intuitive control of the quadruped motion such as head movement control and individual limb puppeteering.
   Experimental results demonstrate improved motion realism and controllability compared to existing retargeting approaches, highlighting the effectiveness of our framework as a tool for animation and virtual production applications.

\ccsdesc[500]{Computing methodologies~Animation}
\ccsdesc[500]{Computing methodologies~Artificial intelligence}

\printccsdesc   
\end{abstract}  

\section{Introduction}
 
Producing believable animal motion is a core challenge in animation, virtual production, and visual effects. 
While human motion capture has become widely accessible, leveraging it to drive animal characters is still non-trivial. 
In practice, it must either rely on highly trained performers who mimic animal behavior~\cite{perry2014mocap_apes, winquist2024new}, or use standard human motion paired with complex hand-crafted control rigs \cite{mufasa_video}.
Both approaches present limitations: the former is difficult to perform consistently and requires expertise, while the latter may produce unnatural results.

In this work, our goal is to create a system that enables a human to steer quadruped motion, as required in a virtual production use case \cite{mufasa_video}. The quadruped motion must spatially conform to the human motion during locomotion, while still giving the human the ability  to ``puppeteer'' semantic actions, such as sitting, jumping, or the look-at direction of the character. Thus, the final motion should satisfy the following conditions:
1) \textbf{Naturalness}, the resulting motion lies within the distribution of natural quadruped motion, exhibiting plausible gaits, contact patterns, and coordination.
2) \textbf{Semantic consistency}, high-level or abstract motion characteristics such as path, direction, speed, and behavior type are preserved from human motion.

Cross-morphology motion transfer poses substantial challenges due to severe morphological, kinematic, and proportion differences. Even at the level of overall scale and speed, it remains ambiguous what it means for a quadruped to ``follow'' a person while producing realistic motion.  For example, a dog runs much faster than a human, while it can walk at a similar speed to human walking. Moreover, quadrupeds show a more diverse set of gait patterns compared to bipeds.
Importantly, not all motion attributes are affected by this ambiguity to the same extent.
We distinguish two sets of motion attributes: \emph{precise features} that can be directly transferred across the two morphologies (e.g., travel direction or target location), and \emph{imprecise features} (e.g., limb positions or root height) that only provide rough semantic cues rather than exact attributes to follow.
These discrepancies make direct motion transfer an inherently ambiguous task.

Existing approaches typically fall into two categories. 
Optimization-based retargeting methods~\cite{seol2013creature} enforce strong alignment between source and target motion, offering controllability but often lacking realism.
Conversely, data-driven approaches~\cite{zhang2018mode, li2023crossloco, Li2024walkthedog} can produce natural-looking motions, but suffer from the absence of paired data, often lack semantic alignment, and provide limited control over how specific motion features are preserved. 
In this work, we bridge this gap using a quadruped motion prior that generates natural motion while enabling multiple levels of control from human motion.
Our key insight is that a diffusion model trained exclusively on quadruped data learns a strong prior over natural animal motion (the naturalness property).
Furthermore, diffusion models provide flexible methods of controlling the generation (the semantics property). More importantly, those control mechanisms support both direct control via network conditions and looser control with inference-based inpainting (imputation). 
\rev{
This is particularly useful for the ill-posed task of cross-morphology motion transfer 
with precise and imprecise features as the control signals.
}

To this end, we present \emph{Two2Four}, a generative framework for human driven quadruped puppeteering.
It consists of two diffusion stages trained purely on quadruped data, and an inference-time inpainting strategy to achieve direct control over user-specified features of the generated quadruped motion.
By abstracting the human movement into universal features, shared between human and animal, we can train a conditional diffusion model on quadruped data and condition it at test time with human features, such as 2D target location or head direction.
In addition, we leverage an inpainting mechanism at high noise levels to control the quadruped motion generation with motion features that are imprecisely mapped between the two morphologies. 
For instance, a user can select to map the motion of the human's hands (or feet) to the quadruped front legs to achieve direct puppeteering or inpaint the root height to achieve semantic actions such as lying, sitting, or jumping. 
This flexible mapping of approximately mapped features stands in contrast to traditional optimization-based or purely data-driven methods, which often require rigid objectives or fixed training distributions. 
Furthermore, we show that splitting the generation process into two stages---first a trajectory stage for a subset of joints and a second stage for full-body motion---improves motion alignment and enables more effective inpainting when trained on a small dataset.
This design makes our system robust, even when operating with out-of-domain control signals.
Our method achieves smooth long-horizon motion generation by autoregressively rolling out the generation.

We evaluate our method on a diverse range of actions. Our experiments show successful transfer of locomotion behaviors such as walking, running, and jumping, as well as finer control such as look-at control and individual limb puppeteering. We further show examples of transferring sitting actions or mapping a human sitting to a quadruped lying down.
Our method achieves more ease and flexibility in control, while resulting in more realistic motion and less foot sliding compared to baseline state-of-the-art methods.
In summary, our contributions are as follows:
\begin{itemize}
    \item A generative framework for quadruped motion synthesis with human driven attributes, trained purely on quadruped data.
    \item A two-stage diffusion model that enables inpainting when dealing with small datasets and improves motion quality.
    \item A hybrid formulation for robustness to out-of-domain data: conditioning on universally shared (sparse) features and an inpainting strategy for imprecisely mapped features.
\end{itemize}

\section{Related Work}

\begin{figure*}
     \centering
     \includegraphics[width=\linewidth]{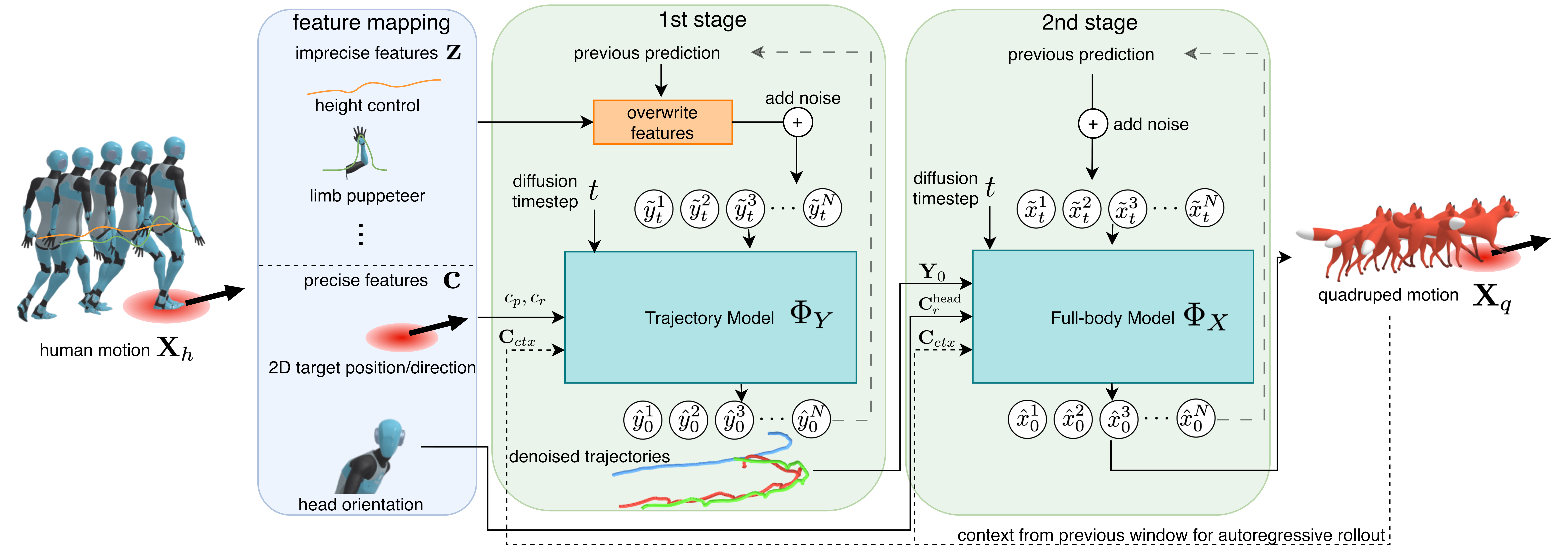}
     \caption{Two2Four is a two-stage diffusion-based pipeline for transferring human motions to quadruped movements. The first stage denoises quadruped trajectories on a subset of joints and enables controllability via conditioning and inpainting. The second stage maps the subset of trajectories to full quadruped motion.   At test time, human motion features are mapped to quadruped motion features and drive the two-stage model via directs conditions for precise features or inpainting for more roughly mapped motion features. For long horizon control, the motion is autoregressively generated by using context frames from the previous motion window. }
     \label{fig:pipeline}
\end{figure*}

Early work in human-driven puppeteering of virtual creatures \cite{Gleicher1998} used rigid feature correspondences for motion transfer, demanding significant manual effort and limiting runtime flexibility.
Recently, learning-based approaches \cite{Zhao2024pose2motion} have relied on inferring correlations across morphologies, improving motion quality but offering less direct control.
In contrast, robustness of generative models \cite{lugmayr2022repaint, mdm} to out-of-domain conditions suggests their potential to enable human-to-quadruped motion transfer with precise actor control without complex feature engineering.
Consequently, the remaining section reviews non-parametric motion transfer, neural motion retargeting, and generative approaches.

\subsection{Non-parametric Motion Transfer}

Early approaches to cross-morphology control relied heavily on explicit pose-level mathematical mappings.
\cite{seol2013creature} proposed building correspondence between human and target creature poses using manually defined optimization goals, followed by per-pose replacement during the puppetry phase.
\cite{rhodin2014interactive} similarly learned pose-mapping from a small user-assembled dataset and allowed transferring motion directly to unrigged target meshes.
More recently, \cite{Doc_Grandia_2023} use differentiable optimal control to transfer animal motion to different robot skeletons of varying proportions.
\cite{agata2025motion} build a transport plan between two motion sources by aligning the pairwise distance matrix among the source frames to that in the target frames.
Motion2Motion~\cite{chen2025motion2motion} extract overlapping patches from source motion and project them into the target skeleton space using sparse mapping, followed by per-patch retrieval and blending, to give the final motion.
Although these methods demonstrate realistic results for their specific applications, for generalized motion they exhibit reduced naturalness due to maintaining rigid correspondences that do not generalize to new contexts.

\subsection{Neural Motion Retargeting}
In the context of intra-morphology retargeting, human-to-human retargeting has flourished with the application of deep learning to large datasets with diverse skeletal structure and body sizes \cite{Lim2019, Aberman2020, lee2023same}.
Early work from Villegas et al.~\cite{Villegas2018} used adversarial cyclic training for unsupervised motion retargeting.
\cite{Aberman2020} introduced using a skeletal convolution network to unify diverse morphologies to a simplified primal skeleton, with \cite{lee2023same} extending this to learn skeleton-agnostic motion embeddings.
Conversely, little research has focused on creature-to-creature retargeting, with the notable exception of \cite{Zhao2024pose2motion}. However, this approach assumes similar motion priors across source and target creatures and requires a specific pose prior for the target skeleton.
Ultimately, these methods fall short when applied to different morphologies due to the substantial domain gap between human and creature motion data.

Regarding cross-morphology motion transfer, \cite{Li2023ace} introduced adversarial losses to facilitate retargeting from humans to different morphologies.
\cite{Egan2024} leveraged shared codebooks learned from manually designed rule-based retargeting, whereas Li et al.~\cite{Li2024walkthedog} proposed VQ-PAE to learn common motion representations in a fully unsupervised manner for aligning human and quadruped data.
However, their method lacks global information, resulting in significant foot sliding artifacts when employing for our use case. Moreover, in the absence of semantically grounded features to guide the alignment, it may incorrectly map different behaviors across domains.
Li et al.~\cite{li2023crossloco} maximize mutual information between human pose and quadruped robot pose together with a cycle-consistency loss to learn a motion transfer policy, but do not generate natural animal motion.

\subsection{Generative Motion}
MDM~\cite{mdm} and EDGE~\cite{Tseng2023} first implemented diffusion models~\cite{Ho2020} for human motion synthesis from text and music, respectively. For inference-time control, MDM and \cite{mu2025stablemotion} explore inpainting, GMD~\cite{Karunratanakul2023gmd} leverage guidance, and DNO~\cite{karunratanakul2024optimizing} show the use of optimization.
\cite{Studer2024, vogeli2025implicit} and \cite{hwang2025motion} employ a two-stage approach with both using a reduced skeletal representation in the first stage. However, they use frame-based Neural IK and temporally-aware diffusion models for the second stage, respectively.
However, none of these methods have demonstrated cross-morphology control or training on smaller datasets such as MANN~\cite{zhang2018mode}.

In contrast, for non-human motion generation, research remains limited due to scarce data.
\cite{li2022ganimator} and \cite{raab2023single} learn generating motion from a single animation clip using adversarial networks and diffusion models, respectively. However, such methods are limited to recomposing the training clip in contrast to generating new behavior.
More recently, OmniMotionGPT~\cite{Yang2024omnimotiongpt} jointly train a motion autoencoder on human and animal datasets, aligning latent representations with text, generating diverse animal motions from text even with limited data. However, this requires text-annotated animal motion datasets.
MAS~\cite{Kapon2024} addresses data scarcity by learning 2D motion priors directly from video and show retargeting to different species. However, it lacks realism especially going from human to quadrupeds.
\cite{gat2025anytop} circumvent data sparsity by training on a diverse set of animals. While they can generalize to new morphologies, they suffer from reduced motion fidelity for any particular skeleton.
\cite{wang2026x} similarly released a large scale motion dataset across 115 species but limited motion data for any particular non-human species. 
Consequently, high-fidelity generalized quadruped motion generation remains underexplored due to  the difficulty in curating large animal motion capture datasets. 

Hence, our work is focused on building a controllable motion generator from a small dataset that is robust to out-of-domain conditions, enabling human actor-driven puppeteering without requiring fine-tuning or complex feature engineering.

\section{Method}
We address the problem of transferring human motion to a quadruped such that the result looks natural and preserves the semantic intent of the source human performance.
Our goal is to generate a corresponding quadruped motion that satisfies naturalness and can transfer different actions such as locomotion, jumping, and sitting. For virtual production, look-at control and direct limb puppeteering are also important.

Our model consists of two-stage diffusion modules, where the first stage captures high-level semantics of motion by only predicting the trajectories of selected joints. 
Then, a second diffusion stage is conditioned on the predicted trajectories and generates the full-body quadruped motion.
At inference time, we map specific features from the human motion to control signals of the diffusion processes via conditions for precisely transferrable features, and via inpainting for imprecise features leaving more flexibility. 
This two-stage decomposition improves motion quality and reduces sliding artifacts compared to a single-stage model~\cite{hwang2025motion}. Additionally, our decomposition increases the effect of inpainting individual trajectories, enabling inpainting as a viable and flexible inference-based control mechanism.
An overview of the full pipeline is depicted in \figref{fig:pipeline}.

Formally, let $\mathbf{X}_{h} = \{\mathbf{x}_h^i\}_{i=1}^{T}$ denote a human motion sequence of length $T$, where each frame $\mathbf{x}_h^i$ represents the human pose. 
Our goal is to generate a corresponding quadruped motion $\mathbf{X}_{q} = \{\mathbf{x}_q^i\}_{i=1}^{T}$, satisfying naturalness and semantic consistency.
Each motion frame is represented as $\mathbf{x}=\{\mathbf{p}, \mathbf{r}\}$ where $\mathbf{p} \in \mathbb{R}^{n\times 3}$ represents the positions and $\mathbf{r} \in \mathbb{R}^{n\times6}$ denotes the 6D rotations~\cite{Zhou2019} of all joints in the skeleton and $n$ denotes the number of joints, which differs for the human and the quadruped.
In the following, we will drop the subscript of the data source $_{q/h}$ for better readability. We always assume quadruped motion unless specifically mentioned.

\subsection{Trajectory Model}
Our first stage aims to predict trajectories of a sparse set of joints $J$, capturing the global structure of quadruped movements.
We represent the trajectory as a sequence $\mathbf{Y}\subset \mathbf{X}$ with $\mathbf{Y}=\{\mathbf{x}^i\}_{i=1}^T\big|_{j \in J~}$, where each frame consists of the selected joints' 3D position and 6D orientation.
The subset includes three joints, namely root, and the two end joints of the front limbs. These joints are selected as they capture the important semantics of the motion, yet are a compact representation, facilitating the inpainting.
We model this distribution using a conditional diffusion model $\Phi_Y$ based on the U-Net model of \cite{cohan2024flexible} that operates in the trajectory space and predicts the clean trajectories $\hat{\mathbf{Y}}_0$. 
Specifically, the model learns to denoise a noisy trajectory $\tilde{\mathbf{Y}}_t$ conditioned on control signals $\mathbf{C}_Y$, parametrized as $\Phi_Y(\tilde{\mathbf{Y}}_{t-1}|\tilde{\mathbf{Y}}_{t}, \mathbf{C}_Y, t)$  where $t$ is the diffusion timestep.
The condition consists of two components. 
First, to enable locomotion control, we specify a target root position $c_p$ and orientation $c_{r}$ to be reached after $T$ frames, which defines the global displacement of the character.
Since this condition is later obtained from the human character, we only maintain the horizontal components of the target position and the yaw component of the rotation. Additionally, we rely on a sparse target signal compared to a dense trajectory condition to give the model more freedom in the generative process, leading to more realism as shown in our ablation in \secref{sec:ablation_traj}.
Second, to ensure temporal continuity when generating long sequences, we condition on a set of preceding context frames $\mathbf{C}_{ctx}=\{\mathbf{x}^{1},..., \mathbf{x}^{n_{c}}\}$, where $n_c$ denotes the number of context frames.
This context enables a smooth transition between consecutive motion windows and mitigates discontinuities that would otherwise arise from independent generation.
Thus, the full condition signal is given by $\mathbf{C}_Y=\{c^T_p,c^T_r,\mathbf{C}_{ctx} \}$.
Both conditions are randomly dropped out during training, with the dropout mask also included in the model input.

The first stage model is trained using a combination of losses:
\begin{equation}
    \mathcal{L} =\mathcal{L}_{recons} + \mathcal{L}_{cond},
\end{equation}
where $\mathcal{L}_{recons}$ is a standard diffusion reconstruction loss between $\mathbf{Y}$ and $\hat{\mathbf{Y}}_0$, and $\mathcal{L}_{cond}$ is a loss for matching the conditions at context frames and final frame. 
\begin{equation}
    \mathcal{L}_{recons} = ||\hat{\mathbf{p}}_0-\mathbf{p}||^2_2 + ||\hat{\mathbf{r}}_0-\mathbf{r}||_2^2,
\end{equation}
\begin{align}
    \mathcal{L}_{cond} &= ||\hat{\mathbf{p}}_{0,xy}^T-{c}_p||^2_2 + ||\hat{\mathbf{r}}^T_{0,xy} - {c}_r||_2^2 \\
    &+ ||\{\hat{\mathbf{x}}_{0}\}_{i=1}^{n_c}|_{j \in J} - \mathbf{C}_{ctx}|_{j \in J} ||_2^2,
\end{align}
where $\hat{\mathbf{p}}^T_{0,xy}$ and $\hat{\mathbf{r}}^T_{0,xy}$ are the 2D position and direction of the root at time T, computed based on the predicted $\hat{\mathbf{Y}}_0$.

\subsection{Full-body Model}
Given the predicted joint trajectories $\hat{\mathbf{Y}}_0$, the second stage aims to generate a full body quadruped motion $\mathbf{X}$ that conforms to the trajectories.
We model this mapping using a conditional diffusion model defined over the full motion space $\Phi_X(\tilde{\mathbf{X}}_{t-1}|\tilde{\mathbf{X}}_{t}, \mathbf{C}_X, t)$. 
The second stage model is conditioned on the subset of joint trajectories $\hat{\mathbf{Y}}_0$ generated by the first stage and the context frames $\mathbf{C}_{ctx}$. 
In addition, head orientations $\mathbf{C}_r^{\text{head}}$ are included in the conditions to enable direct control of the quadruped look-at direction.
Thus the full condition of the second stage model is $\mathbf{C}_X = \{ \hat{\mathbf{Y}}_0, \mathbf{C}_r^{\text{head}}, \mathbf{C}_{ctx} \}$,  and it directly predicts $\hat{\mathbf{X}}_0$.
The model is trained to predict positions and orientations of all joints, along with contact labels of the four limbs.
The loss to train this stage consists of
\begin{equation}
    \mathcal{L} =\mathcal{L}_{recons} +
    \mathcal{L}_{FK} + \mathcal{L}_{sliding} + \mathcal{L}_{cond},
\end{equation}
where $\mathcal{L}_{FK}$ denotes a reconstruction-style loss on the generated movements after applying forward kinematics (FK) with fixed bone length:
\begin{equation}
    \mathcal{L}_{FK}=||\text{FK}(\hat{\mathbf{r}}_0) - \mathbf{p}_0||_2^2.
\end{equation}
    \begin{align}
    \mathcal{L}_{cond}&=||\hat{\mathbf{X}}_0|_{j\in J}-\mathbf{X}|_{j\in J}||_2^2 + ||\hat{\mathbf{r}}_0|_{\text{head}} - \mathbf{C}_r^{\text{head}}||_2^2 \\
    &+ ||\{\hat{\mathbf{x}}_{0}\}_{i=1}^{n_c} - \mathbf{C}_{ctx}||_2^2.
    \end{align}
$\mathcal{L}_{sliding}$ is a standard foot sliding loss based on the predicted contact labels as used in \cite{mdm}.
During training, the trajectory conditions are directly obtained from the ground truth animal dataset.

\subsection{Human to Quadruped Motion Transfer}
\label{sec:method-retargeting}
At inference time, different conditioning and inpainting strategies can be applied to influence the quadruped motion. 
While the trained conditions can be used to map shared features between human and quadruped motion, inpainting enables a looser control by specifying partial abstract or imprecise mappings of motion features between the two domains.
Additionally, due to the inference-based nature of inpainting, this control can be flexibly switched on or off based on the user's preference.

\subsubsection{Precise Feature Mapping with Conditions}
Given the human motion $\mathbf{X}_h$, the target 2D root position and direction of a future frame $T$ are obtained and provided to the first stage model as a condition.
A scaling factor $\gamma$ can be applied on the target 2D position to obtain different action mappings according to the user preference. For instance, scaling up a human walk leads to a running quadruped, while scaling down will keep the quadruped walking.
The resulting joint trajectories from the first stage, along with human head orientation are passed as conditions to the second stage model.
For controlling the head direction $C_r^{\text{head}}$, the human head orientations $C_{r,h}^{\text{head}}$ are mapped to the orientation of quadruped head via a fixed retargeting offset.

\subsubsection{Imprecise Feature Mapping with Inpainting}
\label{sec:inpaint}
To enable additional control, such as limb puppeteering or the height of the quadruped for sitting and jumping, we leverage an inpainting strategy. This overwrites selective components $\mathbf{z}$ of the motion in the first diffusion stage in order to steer the generative process towards these target movements.
At each diffusion timestep $t$, the corresponding components of the noisy sample are overwritten by the noised reference signal.
\begin{equation} 
\tilde{\mathbf{X}}_t|_{\text{z}} = \begin{cases} \sqrt{\bar{\alpha}_t} \mathbf{z} + \sqrt{1-\bar{\alpha}_t}\boldsymbol{\epsilon}, & \text{if } t \geq \tau \\\sqrt{\bar{\alpha}_t} \hat{\mathbf{X}}_{0}|_\text{z}+ \sqrt{1-\bar{\alpha}_t}\boldsymbol{\epsilon}, & \text{if } t < \tau \end{cases}  \quad \text{where } \boldsymbol{\epsilon} \sim \mathcal{N}(\mathbf{0},\mathbf{I}),
\end{equation}
where $|_\text{z}$ denotes the subset of features to be inpainted.
This enforces consistency with the reference signal while allowing the remaining degrees of freedom to be generated by the model.
To balance control and naturalness, the inpainting mask is only active during early diffusion steps and lifted in later diffusion steps $<\tau$.
Thus, rough structural features, which are generated early on in the diffusion process, can be controlled while later diffusion steps maintain realism by matching to the quadruped motion distribution. 
In practice, we find that inpainting the first 80\% denoising steps strikes a good balance between control and naturalness.

In the following, we describe how to transfer different human actions and movements to the desired quadruped movements within our inpainting framework.

\textbf{Sit, lie, and jump:}
Sitting, lying, and jumping behaviors require information about root height.
To transfer these motions, we inpaint the height of the root trajectory during the diffusion process of the trajectory model.
Due to different scales, the target root height is obtained by applying a scale and offset to the human root height,
\begin{equation}
    z_{q}^{root} = \frac{M_{q}-m_{q}}{M_{h}-m_{h}}(z_{h}^{root}-m_{h}) + m_{q}.
    \label{eq:height_scale}
\end{equation}
In the above equation, $m$ and $M$ denote the walking and sitting root height $z^{root}$ in the global frame, respectively. 
For the quadruped, $m_{q}$ and $M_{q}$ are computed from dataset statistic using the median and minimum height, respectively. 
For human, $m_{h}$ and $M_{h}$ can be similarly computed from statistics of the dataset, or set manually.

In practice, different offset values are required depending on the type of source motion. 
For example, human sitting motions may occur either on elevated surfaces (e.g., chairs) or directly on the ground, leading to different relative height offsets.
We leverage this distinction to map different human sitting styles to different quadruped behaviors: sitting on a chair is mapped to a quadruped sitting posture, whereas sitting on the ground is mapped to a lying-down behavior.

For jumping, we replace $M$ to correspond to the maximum heights of human or quadruped jumping.
In practice, we observed that inpainting only the root height provides a weak constraint, as it represents a small subset of the trajectory features and can be ignored by the diffusion model. 
We observe a similar issue when trying to inpaint one feature in a single-stage diffusion, as will be discussed in \secref{sec:ablation}.
Thus, to influence the generation stronger, we additionally inpaint the heights of the remaining joints $z_q^{e}$ in the trajectory model by propagating the root height offset, effectively dragging other joints with the root as it starts to jump,
\begin{equation}
    z^{e}_q=
    \begin{cases}
        \hat{z}_q^{e} + (z_q^{\text{root}}-\hat{z}_q^{\text{root}}) & \text{if } z_q^{\text{root}}>m_q + \varepsilon \\
        \hat{z}_q^{e} & \text{if } z_q^{\text{root}} < m_q + \varepsilon 
    \end{cases}~,
\end{equation}
where $\hat{z}$ denotes the predicted height from the previous diffusion step and $\varepsilon$ is a tolerance threshold.
To accommodate the discrepancy in jumping characteristics, quadrupeds typically cover greater horizontal distances compared to their size, we scale the 2D target position by a constant factor $\gamma>1$ during jump inpainting.

\textbf{Limb puppeteer:}
To enable user-guided manipulation of the quadruped’s limbs using human hand motion, we inpaint the heights of the corresponding limb joints.
The inpainting process operates on the first-stage joint trajectories, where the vertical positions of the quadruped’s limbs are adjusted based on the human hand positions. 
Specifically, the human hand heights are scaled and shifted to match the quadruped’s limb range, producing target heights,
\begin{equation}
    z^{e}_q=sz_h^{e}+b.
\end{equation}
By inpainting only the height dimension, the diffusion model maintains natural joint trajectories while allowing the user to control vertical motion, effectively puppeteering the quadruped’s limbs in a manner consistent with the human demonstration.

\subsection{Implementation Details}
All positions and orientations are expressed in the coordinate frame of the root projected on the ground of the first context frame.
Furthermore, all channels are normalized with a per-joint normalization factor based on the mean and variance of that channel for the specific joint. We find this to improve inpainting compared to normalizing all joints with a single factor.

Our pipeline works in an autoregressive manner to generate long motion sequences.
At the start of a motion, if no previous frames exist, context frames are selected from a default pose (i.e., standing, sitting). Otherwise the last $n_c$ frames of the previously generated window are used as the context.
After generation, the overlapping context frames are linearly blended between the two consecutive windows. Due to the shared context, no complex blending during the diffusion process \cite{Tseng2023} is required.

Both diffusion stages are trained purely on quadruped motion capture data from~\cite{zhang2018mode}, containing various actions (walk, run, jump, sit, lie down). This includes about 40 minutes of quadruped motion, which is small compared to datasets typically used for generative models, which range from tens to hundreds of hours \cite{humanML3D,rempe2026kimodo}.
The model architecture is a temporal U-Net as in \cite{cohan2024flexible} that operates on 32-frame windows with 2-frame context conditioning for temporal coherence and we keep the remaining architecture unchanged.
Choosing a long window size may result in ambiguous behaviors as the human may turn left or right and come to the same target point, whereas a very small window makes the generation more restrictive due to context frames.

We use 1000 diffusion steps during training and DDIM sampling \cite{song2020denoising} with 25 steps during inference.
\rev{
 The combined inference time of both diffusion stages per 32-frame window is 203\,ms on a single RTX 4090. 
 In an offline setting, this allows a 10-second sequence to be generated in approximately 4 seconds. For online applications, the system can generate motions in real-time with a latency of ~0.7\,s, consisting of 0.5\,s of future-target context and 0.2\,s of computation time.
}
The models are optimized with AdamW optimizer~\cite{loshchilov2019decoupled} with adaptive learning rate and batch size of 128.
The training takes roughly 15 and 21 hours for the first and second stage models on a single Nvidia GeForce RTX 3090 GPU.

\rev{
The mapping parameters for inpainting are summarized in \tabref{tab:params}.
Since these parameters are applied only at inference time, tuning and iterating on their values is fast in practice.
Once a suitable set of parameters has been selected, it can be fixed and reused, enabling users to simply choose a behavior mode (e.g., sitting, jumping, or puppeteering) and directly apply the corresponding preset without further adjustment.
}

\begin{table}[h]
    \caption{Mapping parameters for different motion categories.}
    \centering
    \begin{tabular}{lcccc}
        \toprule
        Category & Parameter & Value & Parameter & Value \\
        \midrule
        {General} 
            & $m_q$ & 0.47 
            & $m_h$ & 0.92 \\
        {Jump} 
            & $M_q$ & 1.24 
            & $M_h$ & 1.32 \\
        {Sit \& Lie} 
            & $M_q$ & 0.05 
            & $M_h$ & 0.46 \\
        {Puppeteer} 
            & $s$ & 0.77 
            & $b$ & -0.68 \\
        \bottomrule
    \end{tabular}
    \label{tab:params}
\end{table}
\section{Results}
In this section, we present several experiments using our proposed approach for transferring various human motion to quadruped motions.
The human motions are obtained either from the Human-loco dataset~\cite{Starke2019nsm}, or from our captured motion videos which are converted to skeleton format using a markerless motion capture system \cite{synth2track}.
We refer readers to the supplementary video for a more comprehensive demonstration of the results.

\subsection{Locomotion and Head Control}

\begin{figure}
    \centering
    \includegraphics[width=\linewidth]{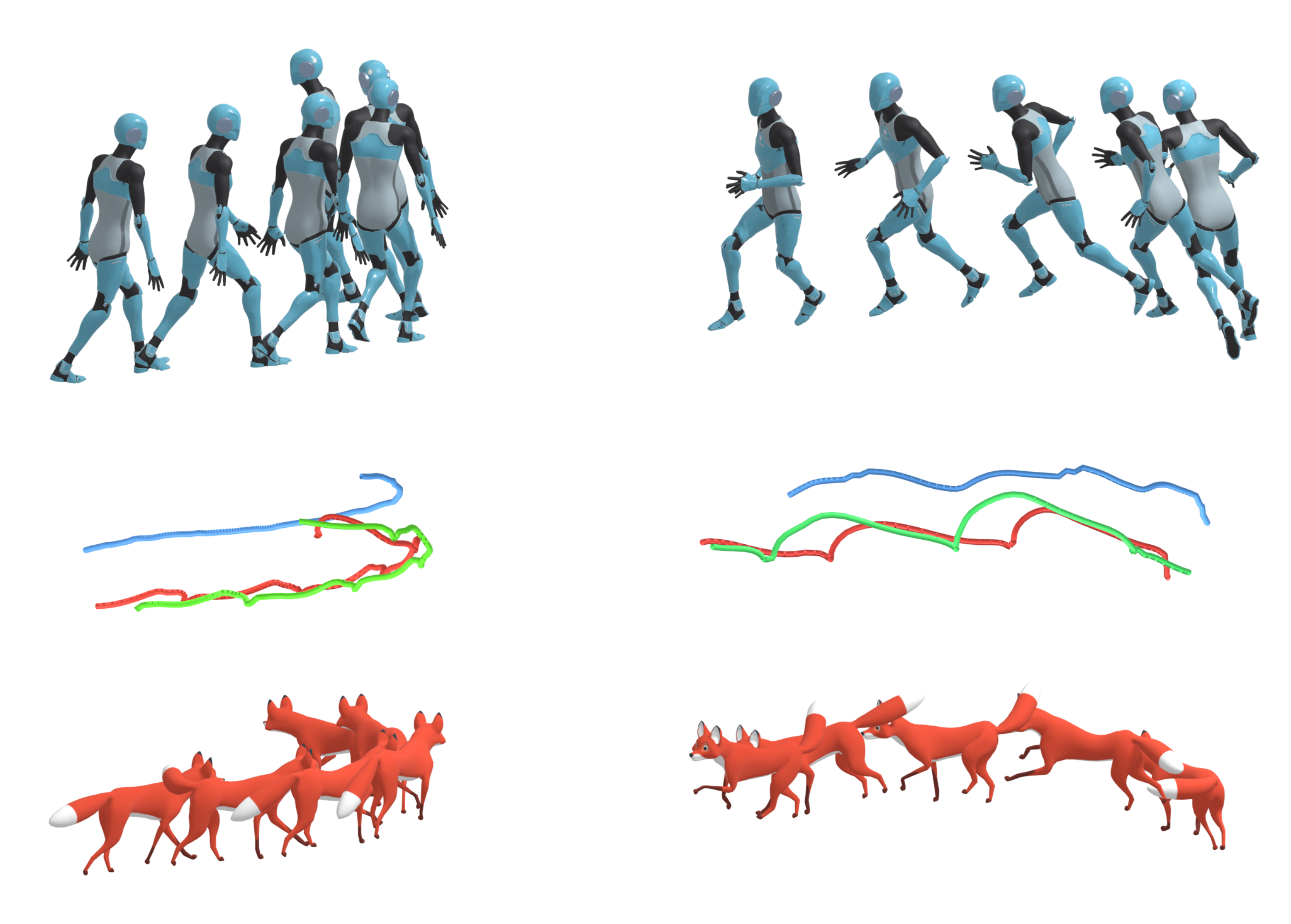}
    \caption{Retargeting Locomotion. Top: Human source motion. Middle: Key trajectories generated by trajectory model. Bottom: Quadruped motion generated by full-body model.}
    \label{fig:results_locomotion}
\end{figure}

Our two-stage model can map human locomotion to quadruped motions by direct conditioning.
\figref{fig:results_locomotion} shows two examples where human walking and running motions are transferred to quadruped walking and running, respectively. 
A scaling factor of $\gamma=0.8$ is used to adjust the 2D target location from human data. By varying this scaling factor, one can achieve different behaviors for the quadruped (e.g., different gait patterns), or make the motion compatible for various quadruped sizes as shown in our supplementary video.

\begin{figure}[t]
    \centering
        \includegraphics[width=\linewidth]{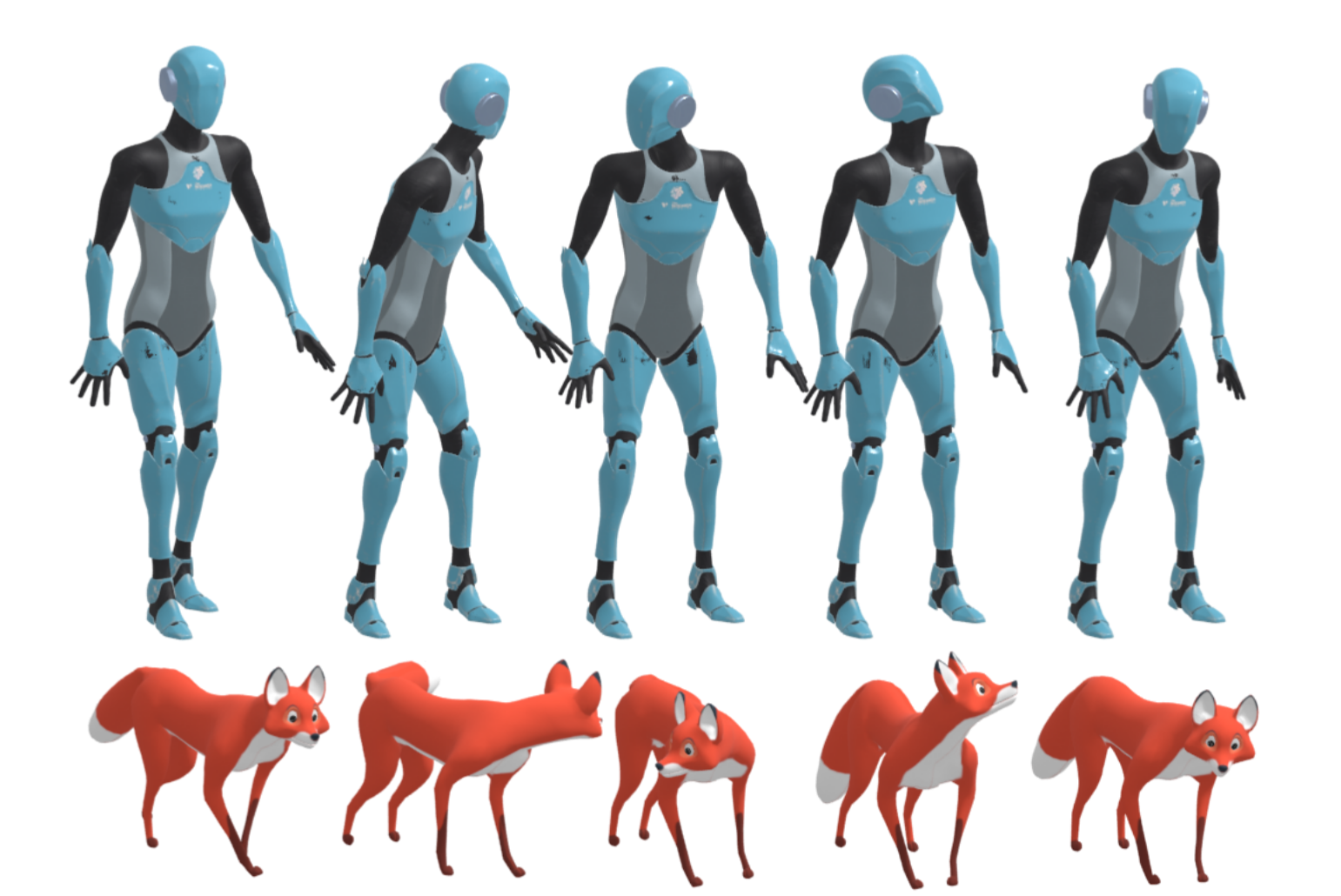}
        \includegraphics[width=\linewidth]{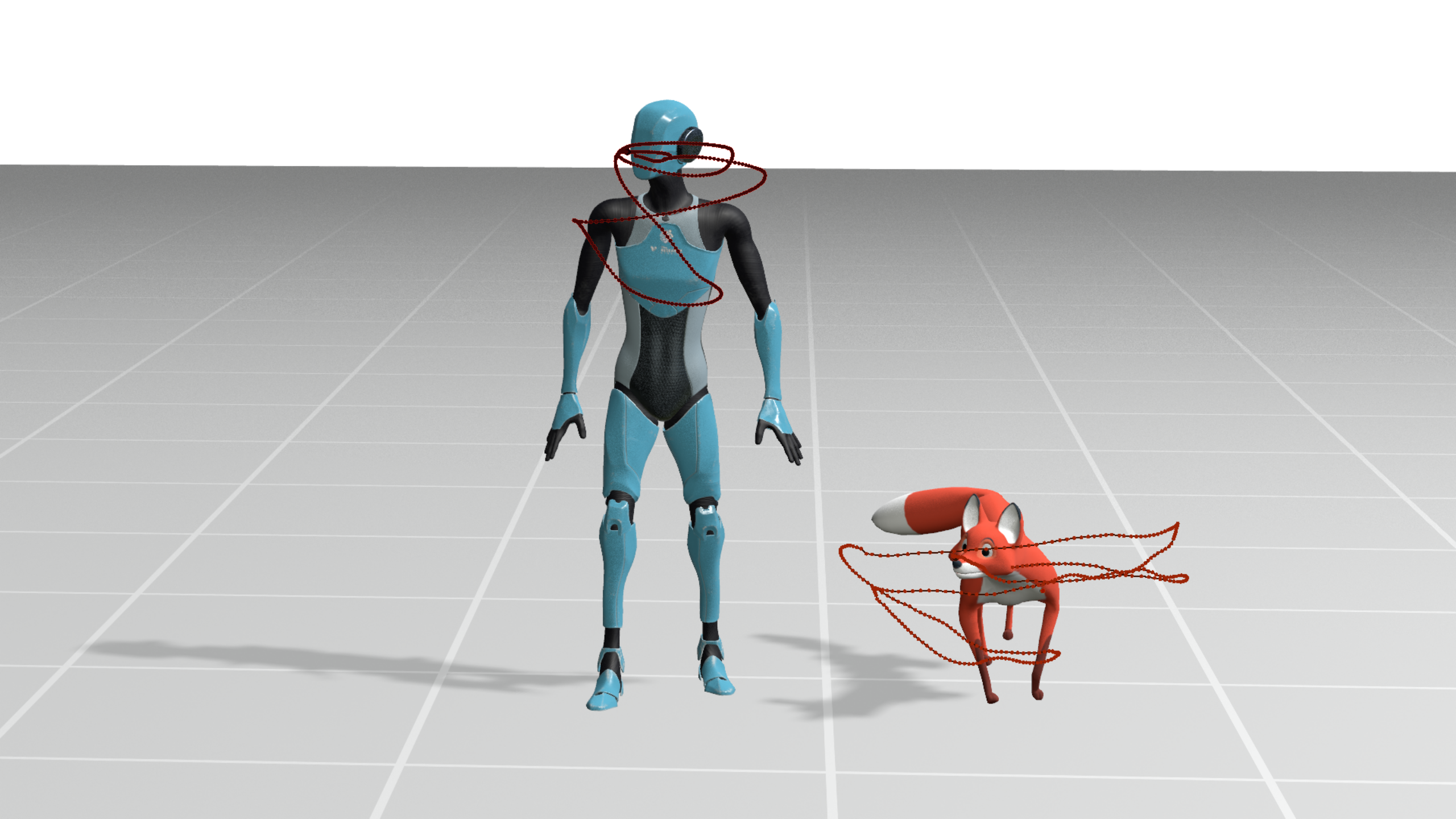}
    \caption{Our second stage full-body model provides direct control of the quadruped's head movements, given the actor's head orientation. Top images show a few snapshots at specific times, bottom shows a full trajectory of the head joint. }
    \label{fig:results_head}
\end{figure}

Look-at control is particularly important in movie production \cite{mufasa_video} when acting out the interactions between characters or the scene.
As shown in the supplementary video, in the absence of explicit conditioning, the quadruped tends to look toward the ground, reflecting the dominant look-at direction in the training data.
Despite the discrepancy between human and quadruped motion domains, the model successfully adapts these signals, producing motions that follow the intended look-at direction, as illustrated in \figref{fig:results_head}.

\subsection{Jumping, Sitting, and Lying}
\begin{figure}[t]
    \centering
    \includegraphics[width=0.5\textwidth]{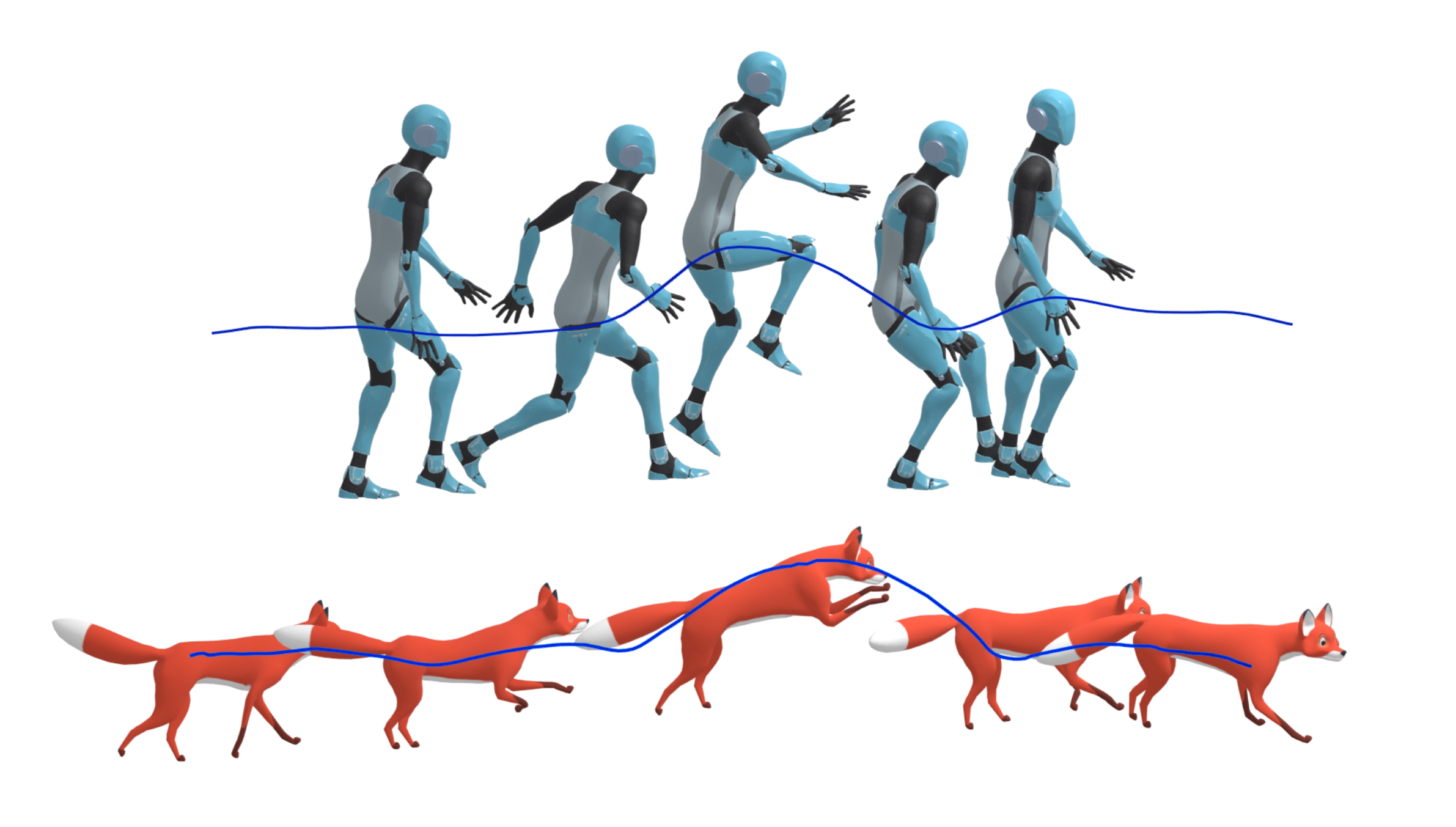}
    \caption{An example jumping motion being produced via inpainting in the trajectory model. 
    }
    \label{fig:jump_demonstration}
\end{figure}

\begin{figure}
    \centering
    \includegraphics[width=0.5\textwidth]{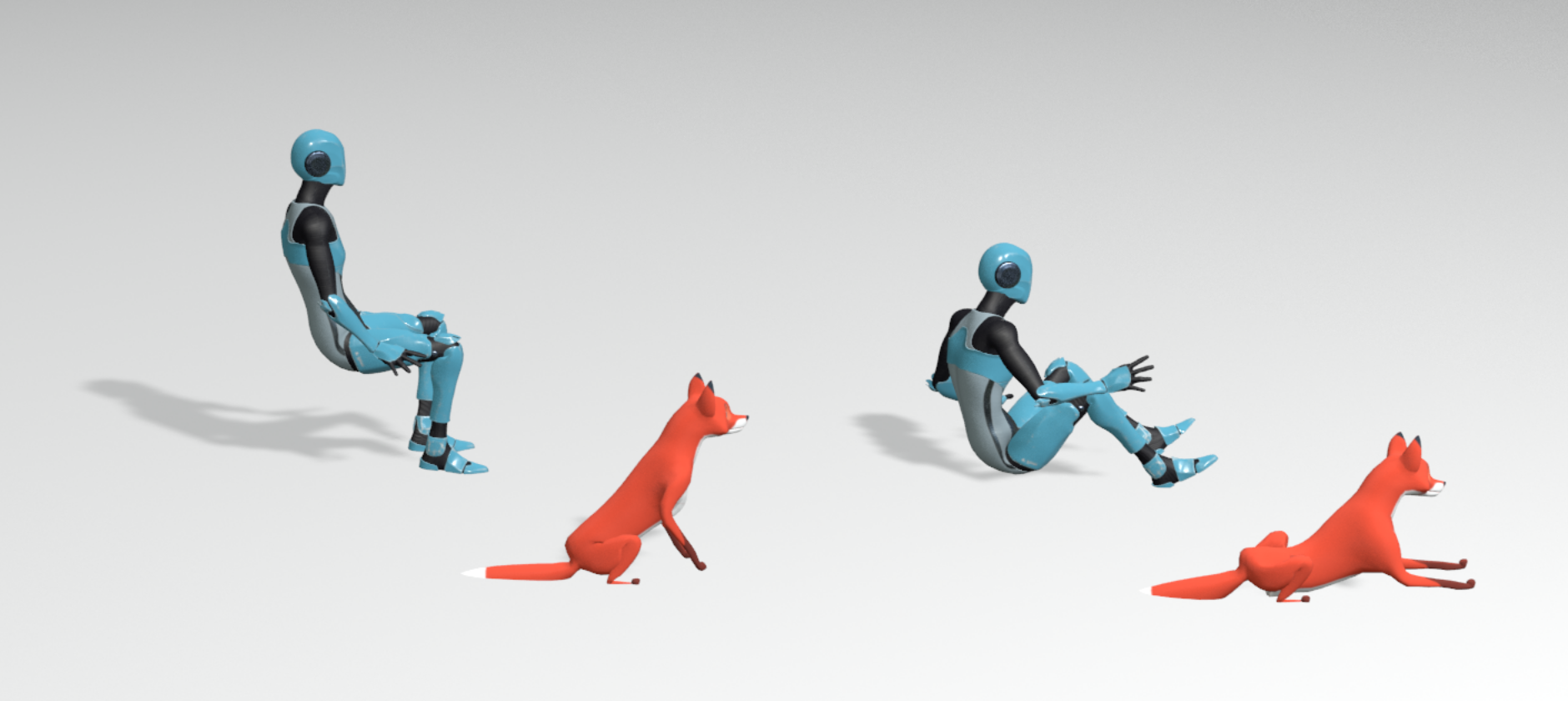}
    \caption{An example of our sitting inpainting strategy: ``sitting on a chair'' gets mapped to quadruped sit (left) and ``sitting on the ground'' is mapped to lying down (right).}
    \label{fig:sit_demonstration}
\end{figure}

Our method leverages inpainting-based control to transfer high-level human actions to quadruped motion.
Using the jump inpainting strategy along with $\gamma=1.5$, our method enables converting human jumping motions to corresponding quadruped jumps, as depicted in \figref{fig:jump_demonstration}.
For sitting and lying experiments, we employ the scaling values detailed in \tabref{tab:params}. This maps human sitting on a chair to quadruped sitting, and human sitting on the ground to quadruped lying, as shown in \figref{fig:sit_demonstration}. By modifying this scale, one can also map human sitting on ground to quadruped sitting if desired.
We note that jumping or sitting inpainting strategies can remain active during the locomotion phase without disrupting the motion.

\subsection{Limb Puppeteering}
\begin{figure}
    \centering
    \includegraphics[width=\linewidth]{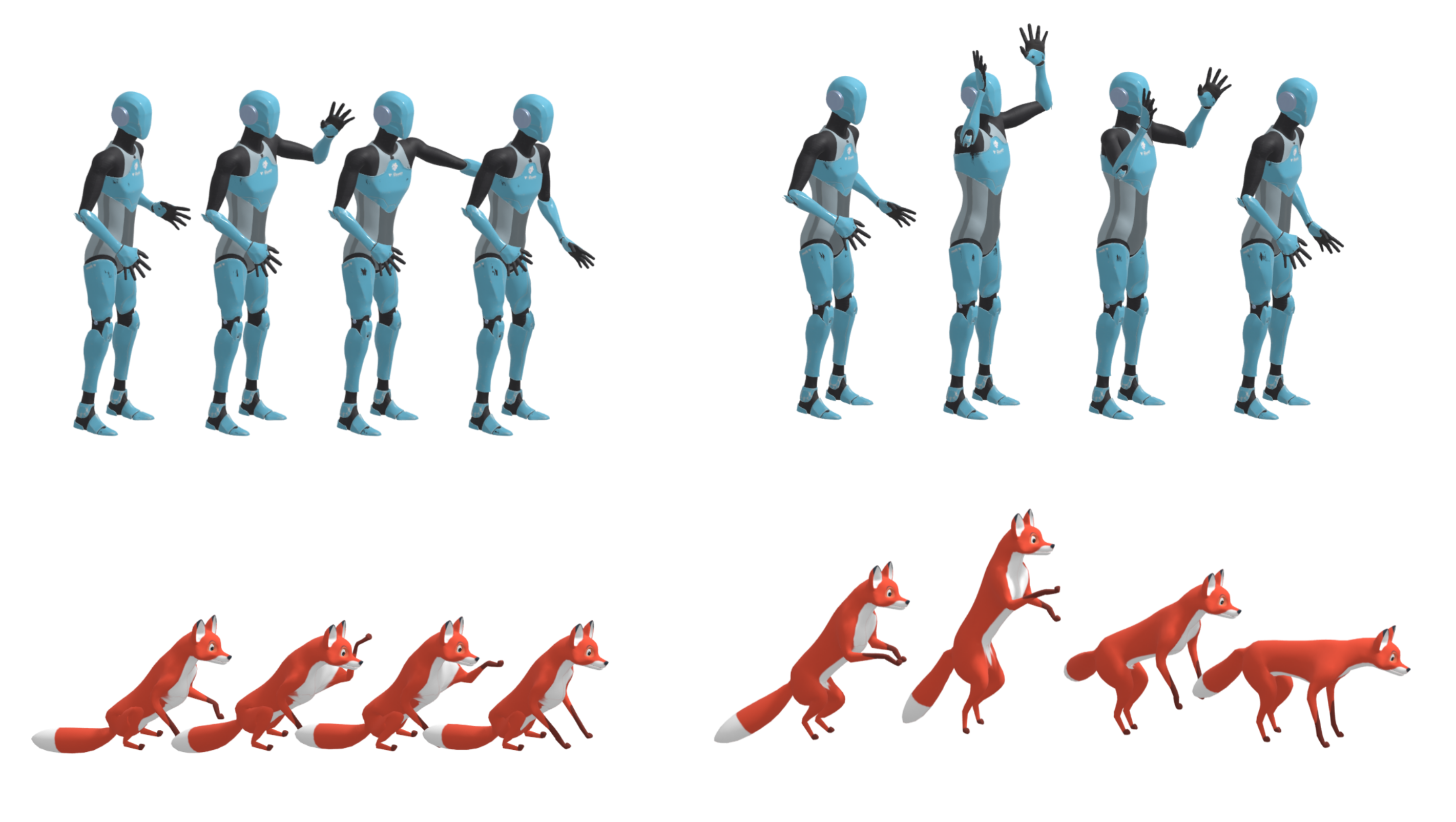}
    \caption{Examples of limb puppeteering via inpainting. 
    }
    \label{fig:results_puppeteer}
\end{figure}

As shown in \figref{fig:results_puppeteer} and our supplementary video, the limb puppeteering inpainting can be used to faithfully control the front legs of the quadruped. 
This allows for an intuitive fine-grained control while maintaining realism.
Notably, when the human raises a single arm, the quadruped responds by ``giving a paw'', naturally transitioning into a sitting posture for stability. When both arms are raised, the quadruped briefly rears up, demonstrating coherent and context-aware whole-body adaptation.
As additionally shown in the supplementary video, the mapping can also be done from another joint source, such as the legs. 

\subsection{Baseline Comparison}
In this section, we compare our method with state-of-the-art baselines most relevant to human-to-quadruped motion transfer.

Qualitatively, we compare our method with \emph{Dog Code}~\cite{Egan2024}, \emph{Walk-the-Dog}~\cite{Li2024walkthedog} and \emph{Motion2Motion}~\cite{chen2025motion2motion}, as summarized in \tabref{tab:qualitative_comparison}. 
We asked two professional artists to rate the realism of the resulting motion for all approaches. 
\textbf{Realism (Local)} refers to motion quality when viewed in local space (zero root movement). 
\textbf{Realism (Global)} refers to motion quality when considering the root movement in global space.
\rev{
\textbf{Semantic Alignment} denotes whether high-level motion intent is preserved from the source motion, e.g., jumping mapped to jumping and walking mapped to walking.
}
\textbf{Direct control} refers to the ability to control specific features of the dog motion from human motion, such as head orientation or direct limb puppeteering.

\emph{Dog Code} shows example of intuitive human to quadruped motion transfer with direct control, however, as seen in their supplementary video (2:10-2:45), the generated dog motion shows issues regarding realism and foot sliding during locomotion.

\rev{
\emph{Motion2Motion} lacks mechanisms to include global information, leading to foot sliding artifacts. Furthermore, its evaluation is limited to short 2-second motion clips. When applied to human-to-quadruped motion transfer on a larger and more diverse motion dataset, it fails to preserve semantic consistency.
}

\emph{Walk-the-Dog} produces realistic motions when viewed in a local frame. 
However, since it is not designed to incorporate global root trajectory information during motion alignment, the resulting motion exhibits noticeable foot sliding when driven with the human root, rendering it unusable for global steering.
Furthermore, because \emph{Walk-the-Dog} relies on completely unsupervised semantic alignment between human and quadruped motions, it can yield misaligned outputs for a given latent encoding. 
For instance, matching a dog jumping motion to a human run, as shown in our supplementary video.

In contrast, \emph{Two2Four} explicitly models global displacement while maintaining naturalness due to the inductive bias of the diffusion model, resulting in overall better realism. It also offers inference-time inpainting techniques that enable direct control of the generated quadruped motion.

\begin{table}[th]
\rev{
    \centering
\caption{Qualitative comparison of SOTA human-to-quadruped motion transfer methods.} 
    \small{
    \begin{tabular}{cccc}
        \toprule
         Approach & \makecell{Realism\\(local / global)}  & \makecell{Semantic\\alignment} & Direct control  \\
         \midrule
         \emph{Walk-the-Dog} & \cmark / \xmark & \xmark & \xmark \\
         \midrule
         \emph{Motion2Motion} & \cmark / \xmark & \xmark & \xmark \\
         \midrule
          \emph{Dog Code} & \xmark / \xmark & \cmark & \cmark \\
          \midrule
          \emph{Ours} & \cmark / \cmark & \cmark & \cmark \\
          \bottomrule
    \end{tabular}
    }
    }
    \label{tab:qualitative_comparison}
\end{table}

In \tabref{tab:quantitative_results}, we quantitatively compare our results with \emph{Walk-the-Dog} on a subset of \emph{HumanLoco} \cite{Starke2019nsm} dataset.
\emph{Walk-the-Dog} copies the 2D root position and yaw angle from human trajectory, while getting the height, roll, pitch and other joint rotations from the motion matching results.
For a fair comparison, we use the scaling factor $\gamma=1$ for this evaluation.
\rev{
We evaluate the motion quality using Fréchet Inception Distance (FID)~\cite{heusel2017gans} computed from a variational auto-encoder (VAE) trained on the same dataset.
}
Our method shows better motion quality as measured by FID and significantly less foot sliding.
When driving our framework with dog motions (reconstruction), FID score and foot sliding values remain close to ground truth data, as reported in \tabref{tab:quantitative_results}.
The large gap in FID values between retarget (human-to-quadruped) and reconstruct (quadruped-to-quadruped) is due to the different distributions of the datasets used for deriving the motion. 

\begin{table}[ht]
    \centering
    \caption{Quantitative analysis of the motion quality. Our method shows better FID and foot sliding compared to Walk-the-Dog.}
    \begin{tabular}{lccc}
            \toprule
        Model & FID $\downarrow$ & Foot sliding $\downarrow$ \\
                \midrule
        GT validation & 0.6838 & 0.055 \\
        Ours (Reconstruct) & 0.9668 & 0.059 \\
        \midrule
        Walk-the-dog (Retarget) & 14.734 & 0.626 \\
        Ours (Retarget)  & \textbf{8.185} & \textbf{0.076} \\
        
        \bottomrule
    \end{tabular}\label{tab:quantitative_results}
\end{table}

\subsection{Ablations}
\label{sec:ablation}
In the following, we ablate the key design choices of our approach, such as the two-stage nature, the sparsity of the locomotion target, and the inpainting strategy.

\subsubsection{One- vs.\ Two-stage Diffusion}
\begin{figure}[t]
    \centering
    \includegraphics[width=\linewidth]{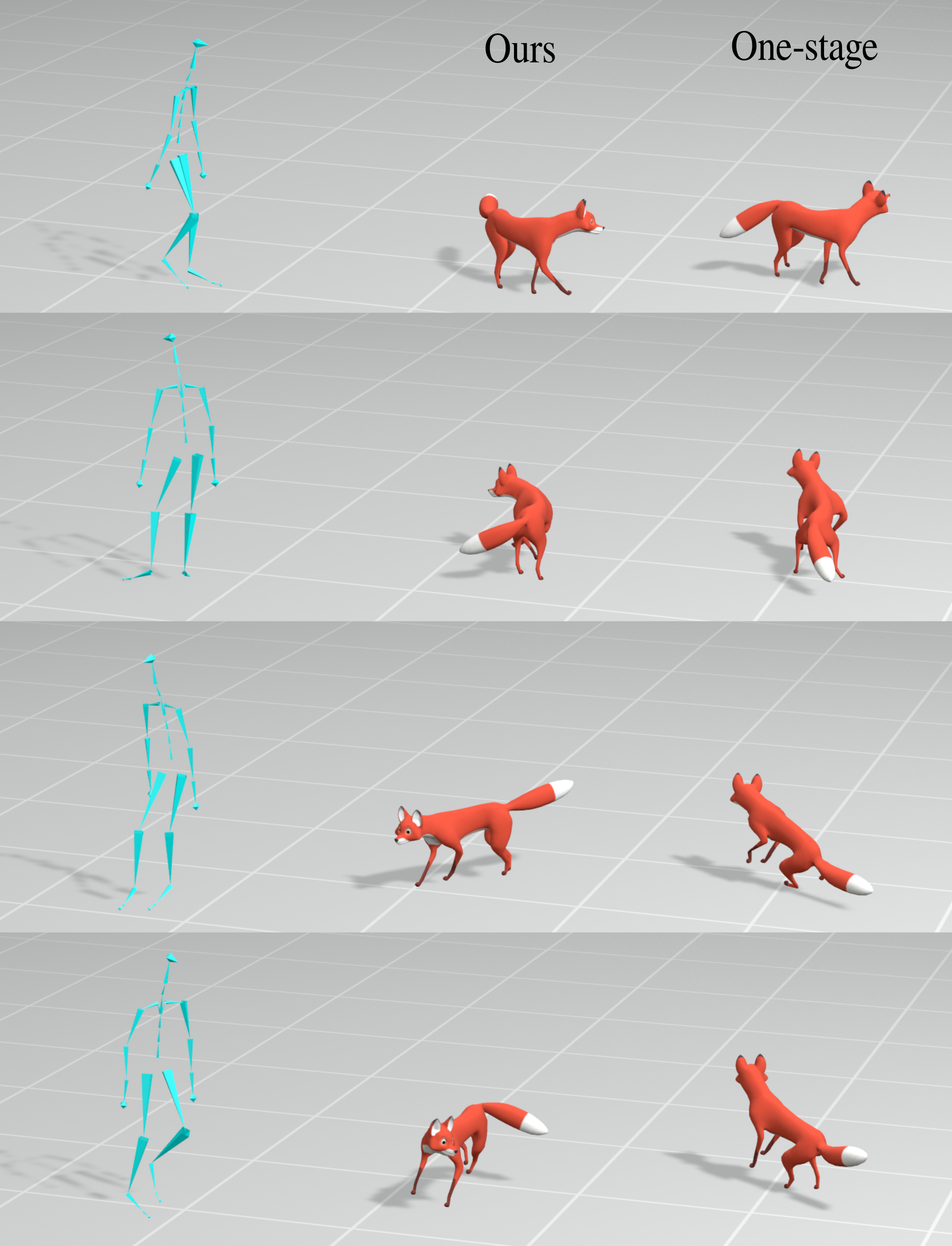}
    \caption{Our result follows the root direction of the human better while the single stage model fails to turn in a similar manner.}
    \label{fig:ablation-1stage}
\end{figure}

We compare our method with a single stage diffusion model (1SD) trained to predict full quadruped motion directly from 2D target position and direction.
We observe that in challenging locomotion scenarios, the two-stage model shows better alignment with human motion.
An example of the alignment issue is depicted in \figref{fig:ablation-1stage}. When the human performs sharp turns, our model can consistently exhibit similar turn behavior, while the single stage model fails to keep the original turn direction. 

\begin{figure}
    \centering
    \includegraphics[width=\linewidth]{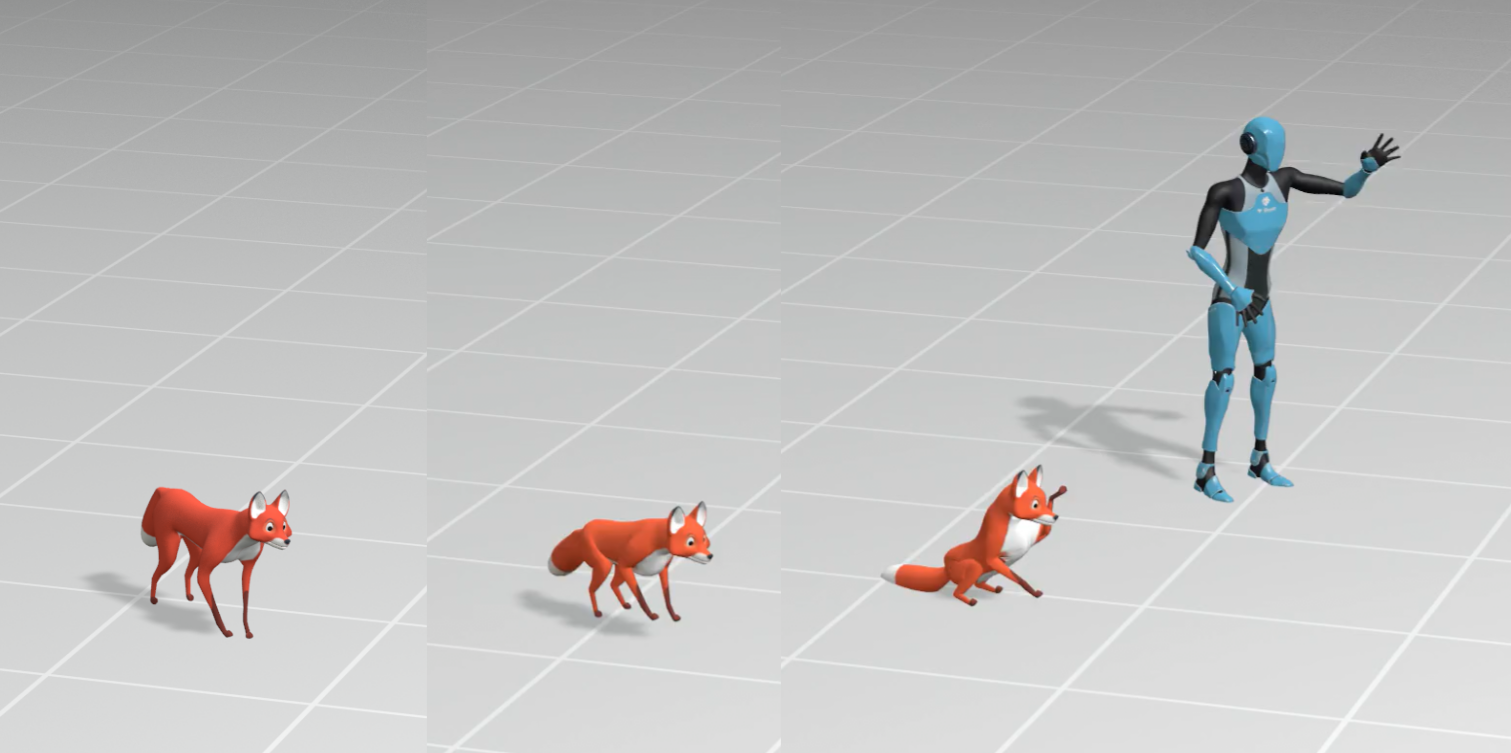}
    \caption{Inpainting front limb trajectories is effective when diffusion model is predicting fewer signals. Single stage diffusion 1SD (left) and two-stage model predicting five joints (middle) fail to follow human hand motion, while our two-stage model predicting three joints (right) can successfully be directed.}
    \label{fig:ablation-puppeteer}
\end{figure}

More importantly, we observe that inpainting individual joints is less effective for 1SD, as shown in \figref{fig:ablation-puppeteer}.
A similar effect is observed when the trajectory model is trained to predict five joints instead of three.
We attribute this to diminishing signal-to-noise ratio when the diffusion model has to denoise more data.

\subsubsection{Sparse Goal vs.\ Full Trajectory}
\label{sec:ablation_traj}
\begin{figure}
    \centering
    \includegraphics[width=\linewidth]{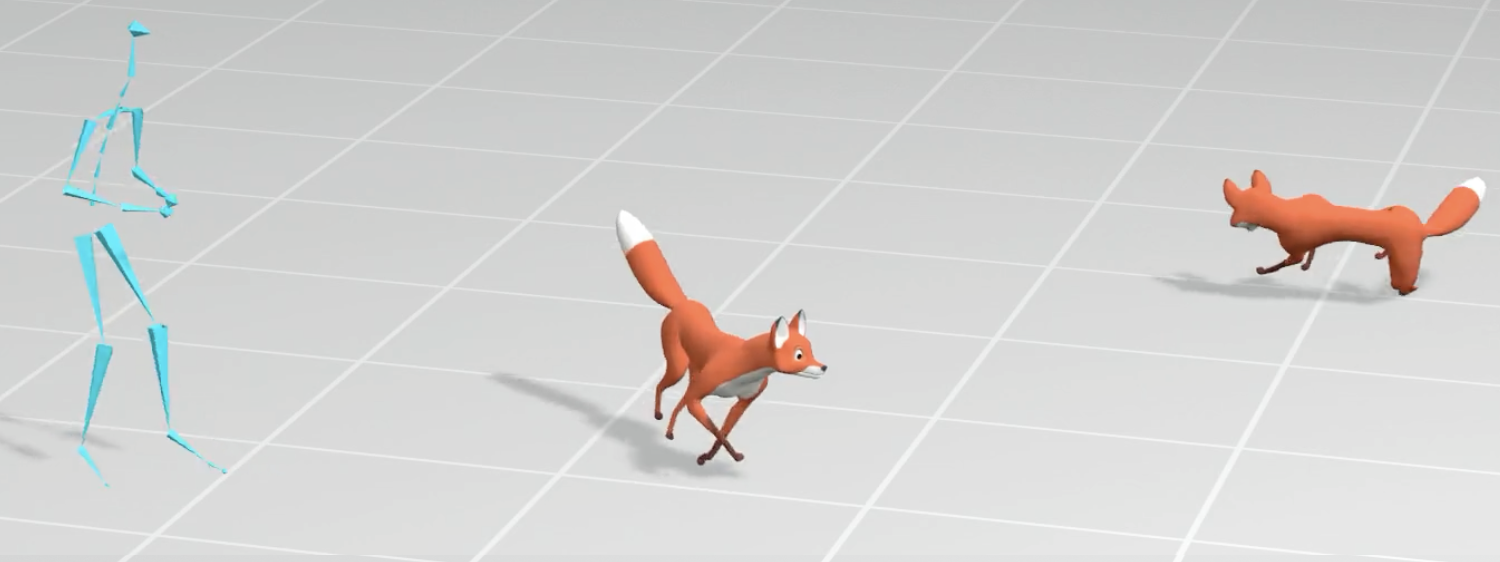}
    \caption{A model trained on full trajectory as condition (right) leads to artifacts compared two our sparse target condition (left).}
    \label{fig:ablation-dense}
\end{figure}

In the following, we ablate choosing a sparse root target at frame $T$ against having the full root trajectory as condition to the trajectory model.
For simple motion sequences, both approaches perform comparably when transferring human motion to quadrupeds. However, for more challenging cases, such as sharp in-place turns, the full conditioned model struggles to follow the motion and produces noticeable artifacts, as shown in \figref{fig:ablation-dense}. In contrast, the sparse-target model remains stable and follows the motion even in such challenging cases.
We attribute this to the condition signal getting too far out-of-distribution for the model to recover, a problem that is more prominent for models trained on small datasets such as ours.

\subsubsection{Inpainting Schedule}
\label{sec:ablation_inpaint}
To account for the imprecise mapping of some motion features between human and quadruped, a key component in our inpainting control is to  apply it only until a fixed percentage of diffusion time steps, in our case 80\%.
As shown in the supplementary video, varying this threshold for inpainting leads to different results.
Fewer inpainting steps give more flexibility to the diffusion model, but may weaken the adherence to the intended control signal.
Conversely, inpainting all diffusion steps can introduce artifacts and reduce naturalness. For instance, in jumping motions, full inpainting leads to excessive acceleration and missing the proper steps for the jump. In contrast, in our setting the quadruped takes an extra step to prepare for the jump and the jump is smoother.

\section{Discussion}

Compared to optimization-based frameworks with rigid mappings, the diffusion process enables more flexible mapping to the target distribution, as precise features can be concretely mapped as conditions and imprecisely mapped motion features can be weakly transferred via inpainting.
Crucially, we condition the generator on a sparse goal position rather than a full body trajectory.
Because quadrupeds cannot turn in place like humans, over-constraining the trajectory leads to unnatural movement; sparse conditioning avoids this, allowing for realistic quadrupedal motion to the human target location.
In contrast, the head orientation trajectory can be seamlessly transferred without artifacts. Hence, we condition our second stage on ground-truth quadruped head orientation during training and use the corresponding human look-at signal during inference.

Moreover, at early diffusion steps, the out-of-domain inpainted features, such as root height, are sufficiently distorted by the large added noise to resemble the in-distribution features.
Hence, the generative model synthesizes motion conforming to these imprecise features while maintaining realism as long as the inpainting is dropped at lower noise levels.
Besides inpainting, other forms of inference-based influence, such as guidance~\cite{Karunratanakul2023gmd} and noise optimization~\cite{karunratanakul2024optimizing}, have been explored for motion generation. However, compared to our inpainting strategy, they can come with a larger computational cost which is more challenging for a real-time use case, such as live puppeteering in virtual production.

When training on small datasets such as MANN~\cite{zhang2018mode}, we observe that increasing the dimensionality of the features, i.e.\ the number of trajectories generated, in the first stage diminishes inpainting effectiveness.
We hypothesize that as more joint trajectories are introduced, the inter-feature correlations increase.
Hence, rather than utilizing the inpainted signal, the model opts to regenerate the trajectory based on the remaining highly correlated features.
Consequently, we observe that the first-stage model generating three trajectories (hip and front feet) outperforms one generating five (hip and all feet), which in turn outperforms a single-stage model generating all joint trajectories.

\rev{
Although our evaluation focuses on dogs, the proposed framework is largely species-agnostic and does not incorporate any dog-specific modeling assumptions. 
Consequently, given sufficient motion data, Two2Four can be adapted to other quadruped species with minimal changes to the overall pipeline. In practice, however, the availability of high-quality quadruped motion datasets remains limited, with canine datasets being the most accessible public resources.
}
\section{Conclusion}
We introduced Two2Four, a two-stage diffusion model along with inference-based inpainting strategies to generate quadruped motion driven by human motion.
We showed that a two-stage diffusion approach results in higher-quality movements than a single-stage model, particularly when driven by out-of-distribution human data.

Our method shows high quality motions and can imitate different actions including locomotion, jumping, sitting, lying, as well as individual limb movements.
However, due to the lack of backward quadruped motion in our dataset, our model struggles when the human is walking backwards for more than a few seconds. 
In this case, the quadruped sits down and shows sliding artifacts. 
Nevertheless, due to the sparse goal condition mechanism and our autoregressive rollout scheme, the quadruped can recover and catch up to the human source after a while. 

In this work, we focus on enabling multiple control modes for human-to-quadruped motion transfer.
Currently, these modes are determined by the user.
Designing an automated system that switches between these modes (for example by setting some thresholds), or finding a unified representation for these modes remains a design challenge to be explored in the future.
Additional future directions could also explore Bézier motion models as in \cite{vogeli2025implicit} to reduce the data dimensionality in order to include more joints in the first stage, thus enabling more user control of the motion, or to allow for artist-friendly editing of the quadruped motion.

\section*{Acknowledgment}
The authors would like to thank Dominik Borer for his technical support and Violaine Fayolle and Doriano van Essen for their artistic support during this project.

\printbibliography   

@String{tog = "ACM TOG"}

@article{perry2014mocap_apes,
  author  = {Tekla S. Perry},
  title   = {Motion Capture Technology Goes Into the Wild for Dawn of the Planet of the Apes},
  journal = {IEEE Spectrum},
  year    = {2014},
  url     = {https://spectrum.ieee.org/motion-capture-technology-goes-into-the-wild-for-dawn-of-the-planet-of-the-apes},
  note    = {Accessed: 2026-04-15}
}

@incollection{winquist2024new,
  title={A New Kingdom: Weta FX Returns to The Planet of The Apes},
  author={Winquist, Erik and Leonhardt, Phillip and Story, Paul and Nowotny, Alex},
  booktitle={ACM SIGGRAPH 2024 Talks},
  pages={1--2},
  year={2024}
}

@inproceedings{
song2020denoising,
title={Denoising Diffusion Implicit Models},
author={Jiaming Song and Chenlin Meng and Stefano Ermon},
booktitle={International Conference on Learning Representations},
year={2021},
}

@inproceedings{chen2025motion2motion,
  title={Motion2motion: Cross-topology motion transfer with sparse correspondence},
  author={Chen, Ling-Hao and Zhang, Yuhong and Yin, Zixin and Dou, Zhiyang and Chen, Xin and Wang, Jingbo and Komura, Taku and Zhang, Lei},
  booktitle={Proceedings of the SIGGRAPH Asia 2025 Conference Papers},
  pages={1--11},
  year={2025}
}

@inproceedings{heusel2017gans,
  title     = {GANs Trained by a Two Time-Scale Update Rule Converge to a Local Nash Equilibrium},
  author    = {Heusel, Martin and Ramsauer, Hubert and Unterthiner, Thomas and Nessler, Bernhard and Hochreiter, Sepp},
  booktitle = {Advances in Neural Information Processing Systems (NeurIPS)},
  year      = {2017}
}

@inproceedings{synth2track,
author = {Buhmann, Jakob and Moore, Douglas L. and Borer, Dominik and Guay, Martin},
title = {Synth2Track Editor for Efficient Match-Animation},
year = {2025},
publisher = {Association for Computing Machinery},
address = {New York, NY, USA},
booktitle = {Proceedings of the Special Interest Group on Computer Graphics and Interactive Techniques Conference Talks},
articleno = {3},
numpages = {3},
location = {
},
series = {SIGGRAPH Talks '25}
}

@article{Doc_Grandia_2023,
author = {Grandia, Ruben and Farshidian, Farbod and Knoop, Espen and Schumacher, Christian and Hutter, Marco and B\"{a}cher, Moritz},
title = {DOC: Differentiable Optimal Control for Retargeting Motions onto Legged Robots},
year = {2023},
issue_date = {August 2023},
publisher = {Association for Computing Machinery},
address = {New York, NY, USA},
volume = {42},
number = {4},
issn = {0730-0301},
doi = {10.1145/3592454},
journal = {ACM Trans. Graph.},
month = jul,
articleno = {96},
numpages = {14}
}

@article{loshchilov2019decoupled,
  title={Decoupled Weight Decay Regularization},
  author={Loshchilov, Ilya and Hutter, Frank},
  journal={International Conference on Learning Representations (ICLR)},
  year={2019}
}

@inproceedings{seol2013creature,
  title={Creature Features: Online Motion Pppetry for Non-Human Characters},
  author={Seol, Yeongho and O'Sullivan, Carol and Lee, Jehee},
  booktitle={Proceedings of the 12th ACM SIGGRAPH/Eurographics Symposium on Computer Animation},
  pages={213--221},
  year={2013},
  publisher={ACM}
}

@article{rhodin2014interactive,
  title={Interactive Motion Mapping for Real-time Character Control},
  author={Rhodin, Helge and Tompkin, James and Kim, Kwang In and Varanasi, Kiran and Seidel, Hans-Peter and Theobalt, Christian},
  journal={Computer Graphics Forum},
  volume={33},
  number={2},
  pages={273--282},
  year={2014},
  publisher={Wiley Online Library}
}

@article{zhang2018mode,
  title={Mode-Adaptive Neural Networks for Quadruped Motion Control},
  author={Zhang, He and Starke, Sebastian and Komura, Taku and Saito, Jun},
  journal={ACM Transactions on Graphics (ToG)},
  volume={37},
  number={4},
  pages={1--11},
  year={2018},
  publisher={ACM New York, NY, USA}
}

@inproceedings{cohan2024flexible,
  title={Flexible motion in-betweening with diffusion models},
  author={Cohan, Setareh and Tevet, Guy and Reda, Daniele and Peng, Xue Bin and van de Panne, Michiel},
  booktitle={ACM SIGGRAPH 2024 conference papers},
  pages={1--9},
  year={2024}
}

@misc{mufasa_video,
  author       = {VFX~express},
  title        = {Mufasa: The Lion King – MPC's Character Lab Brings Taka to Life},
  year         = {2025},
  howpublished = {\url{https://www.youtube.com/watch?v=XCeJu8XeIJ4}},
  note         = {YouTube video, accessed April 10, 2026}
}

@InProceedings{Gleicher1998,
  author = {Michael Gleicher},
  title = {Retargetting Motion to New Characters},
  booktitle = {Proceedings of the 25th Annual Conference on Computer Graphics and Interactive Techniques},
  series = {SIGGRAPH '98},
  year = {1998},
  pages = {33--42},
  publisher = {ACM},
}

@InProceedings{Villegas2018,
  author = {Ruben Villegas and Jimei Yang and Duygu Ceylan and Honglak Lee},
  title = {Neural Kinematic Networks for Unsupervised Motion Retargetting},
  booktitle = {Proceedings of the IEEE Conference on Computer Vision and Pattern Recognition (CVPR)},
  year = {2018},
  pages = {8639--8648},
}

@InProceedings{Lim2019,
  author = {Jongin Lim and Hyung Jin Chang and Jin Young Choi},
  title = {PMnet: Learning of Disentangled Pose and Movement for Unsupervised Motion Retargeting},
  booktitle = {Proceedings of the British Machine Vision Conference (BMVC)},
  volume = {2},
  year = {2019},
  pages = {7}
}

@inproceedings{Li2024walkthedog,
  title={WalktheDog: Cross-Morphology Motion Alignment via Phase Manifolds},
  author={Li, Peizhuo and Starke, Sebastian and Ye, Yuting and Sorkine-Hornung, Olga},
  booktitle={ACM SIGGRAPH 2024 Conference Papers},
  pages={1--10},
  year={2024}
}

@InProceedings{Li2023ace,
  author = {Tianyu Li and Jungdam Won and Alexander Clegg and Jeonghwan Kim and Akshara Rai and Sehoon Ha},
  title = {ACE: Adversarial Correspondence Embedding for Cross Morphology Motion Retargeting from Human to Nonhuman Characters},
  booktitle = {SIGGRAPH Asia 2023 Conference Papers},
  year = {2023},
  pages = {1--11},
}

@Article{Starke2019nsm,
  author = {Sebastian Starke and He Zhang and Taku Komura and Jun Saito},
  title = {Neural State Machine for Character-Scene Interactions},
  journal = {ACM Transactions on Graphics},
  volume = {38},
  number = {6},
  year = {2019},
  Articleno = {209},
  pages = {209:1--209:14},
}

@inproceedings{
mdm,
title={Human Motion Diffusion Model},
author={Guy Tevet and Sigal Raab and Brian Gordon and Yoni Shafir and Daniel Cohen-or and Amit Haim Bermano},
booktitle={The Eleventh International Conference on Learning Representations },
year={2023},
}

@InProceedings{Tseng2023,
  author = {Jonathan Tseng and Rodrigo Castellon and Karen Liu},
  title = {{EDGE}: Editable Dance Generation from Music},
  booktitle = {Proceedings of the IEEE/CVF Conference on Computer Vision and Pattern Recognition},
  year = {2023},
  pages = {448--458},
}

@InProceedings{Ho2020,
  author = {Jonathan Ho and Ajay Jain and Pieter Abbeel},
  title = {Denoising Diffusion Probabilistic Models},
  booktitle = {Advances in Neural Information Processing Systems},
  volume = {33},
  year = {2020},
  pages = {6840--6851}
}

@Misc{Karunratanakul2023gmd,
  author = {Korrawe Karunratanakul and Konpat Preechakul and Supasorn Suwajanakorn and Siyu Tang},
  title = {{GMD}: Controllable Human Motion Synthesis via Guided Diffusion Models},
  journal = {arXiv preprint arXiv:2305.12577},
  year = {2023}
}

@inproceedings{Yang2024omnimotiongpt,
  title={OmniMotionGPT: Animal Motion Generation with Limited Data},
  author={Yang, Zhangsihao and Zhou, Mingyuan and Shan, Mengyi and Wen, Bingbing and Xuan, Ziwei and Hill, Mitch and Bai, Junjie and Qi, Guo-Jun and Wang, Yalin},
  booktitle={Proceedings of the IEEE/CVF Conference on Computer Vision and Pattern Recognition},
  pages={1249--1259},
  year={2024}
}

@InProceedings{Kapon2024,
  author = {Roy Kapon and Guy Tevet and Daniel Cohen-Or and Amit H. Bermano},
  title = {{MAS}: Multi-view Ancestral Sampling for {3D} Motion Generation Using {2D} Diffusion},
  booktitle = {Proceedings of the IEEE/CVF Conference on Computer Vision and Pattern Recognition},
  year = {2024},
  pages = {1965--1974},
}

@InProceedings{Zhao2024pose2motion,
  author = {Qingqing Zhao and Peizhuo Li and Yifan Wang and Olga Sorkine-Hornung and Gordon Wetzstein},
  title = {Pose-to-Motion: Cross-Domain Motion Retargeting with Pose Prior},
  booktitle = {Computer Graphics Forum},
  volume = {43},
  year = {2024},
  pages = {e15170},
  publisher = {Wiley Online Library},
}

@Article{Aberman2020,
  author = {Kfir Aberman and Peizhuo Li and Dani Lischinski and Olga Sorkine-Hornung and Daniel Cohen-Or and Baoquan Chen},
  title = {Skeleton-Aware Networks for Deep Motion Retargeting},
  journal = {ACM Transactions on Graphics},
  volume = {39},
  number = {4},
  year = {2020},
  Articleno = {62},
  pages = {62:1--62:14},
}

@InProceedings{Egan2024,
  author = {Donal Egan and Alberto Jovane and Jan Szkaradek and George Fletcher and Darren Cosker and Rachel McDonnell},
  title = {Dog Code: Human to Quadruped Embodiment Using Shared Codebooks},
  booktitle = {Proceedings of the 17th ACM SIGGRAPH Conference on Motion, Interaction, and Games},
  year = {2024},
  pages = {1--11},
}

@InProceedings{Studer2024,
  author = {Justin Studer and Dhruv Agrawal and Dominik Borer and Seyedmorteza Sadat and Robert W. Sumner and Martin Guay and Jakob Buhmann},
  title = {Factorized Motion Diffusion for Precise and Character-Agnostic Motion In-Betweening},
  booktitle = {Proceedings of the 17th ACM SIGGRAPH Conference on Motion, Interaction, and Games},
  year = {2024},
  pages = {1--10},
}

@InProceedings{Zhou2019,
  author = {Yi Zhou and Connelly Barnes and Jingwan Lu and Jimei Yang and Hao Li},
  title = {On the Continuity of Rotation Representations in Neural Networks},
  booktitle = {Proceedings of the IEEE/CVF Conference on Computer Vision and Pattern Recognition},
  year = {2019},
  pages = {5745--5753},
}

@inproceedings{agata2025motion,
  title={Motion Control via Metric-Aligning Motion Matching},
  author={Agata, Naoki and Igarashi, Takeo},
  booktitle={Proceedings of the Special Interest Group on Computer Graphics and Interactive Techniques Conference Conference Papers},
  pages={1--12},
  year={2025}
}

@inproceedings{lee2023same,
  title={SAME: Skeleton-Agnostic Motion Embedding for Character Animation},
  author={Lee, Sunmin and Kang, Taeho and Park, Jungnam and Lee, Jehee and Won, Jungdam},
  booktitle={SIGGRAPH Asia 2023 Conference Papers},
  pages={1--11},
  year={2023}
}

@inproceedings{vogeli2025implicit,
  title={Implicit B{\'e}zier Motion Model for Precise Spatial and Temporal Control},
  author={V{\"o}geli, Luca and Agrawal, Dhruv and Guay, Martin and Borer, Dominik and Sumner, Robert W and Buhmann, Jakob},
  booktitle={Proceedings of the 2025 18th ACM SIGGRAPH Conference on Motion, Interaction, and Games},
  pages={1--10},
  year={2025}
}

@inproceedings{hwang2025motion,
            title={Motion Synthesis with Sparse and Flexible Keyjoint Control}, 
            author={Inwoo Hwang and Jinseok Bae and Donggeun Lim and Young Min Kim},
            booktitle = {Proceedings of the IEEE/CVF International Conference on Computer Vision (ICCV)},
            year      = {2025},
}

@inproceedings{gat2025anytop,
  title={Anytop: Character animation diffusion with any topology},
  author={Gat, Inbar and Raab, Sigal and Tevet, Guy and Reshef, Yuval and Bermano, Amit Haim and Cohen-Or, Daniel},
  booktitle={Proceedings of the Special Interest Group on Computer Graphics and Interactive Techniques Conference Conference Papers},
  pages={1--10},
  year={2025}
}

@inproceedings{wang2026x,
  title={X-MoGen: Unified Motion Generation Across Humans and Animals},
  author={Wang, Xuan and Ruan, Kai and Qian, Liyang and Zhi, Guo Zhi and Su, Chang and Wang, Gaoang},
  booktitle={Proceedings of the AAAI Conference on Artificial Intelligence},
  volume={40},
  number={12},
  pages={10234--10242},
  year={2026}
}

@inproceedings{li2023crossloco,
  title={CrossLoco: Human Motion Driven Control of Legged Robots via Guided Unsupervised Reinforcement Learning},
  author={Li, Tianyu and Jung, Hyunyoung and Gombolay, Matthew and Cho, Yong and Ha, Sehoon},
  booktitle={The Twelfth International Conference on Learning Representations},
  year={2024}
}

@article{li2022ganimator,
  title={GANimator: Neural Motion Synthesis from a Single Sequence},
  author={Li, Peizhuo and Aberman, Kfir and Zhang, Zihan and Hanocka, Rana and Sorkine-Hornung, Olga},
  journal={ACM Transactions on Graphics (TOG)},
  volume={41},
  number={4},
  pages={1--12},
  year={2022},
  publisher={ACM New York, NY, USA}
}

@inproceedings{raab2023single,
  title={Single Motion Diffusion},
  author={Raab, Sigal and Leibovitch, Inbal and Tevet, Guy and Arar, Moab and Bermano, Amit Haim and Cohen-Or, Daniel},
  booktitle={The Twelfth International Conference on Learning Representations},
  year={2024}
}

@inproceedings{karunratanakul2024optimizing,
  title={Optimizing Diffusion Noise Can Serve as Universal Motion riors},
  author={Karunratanakul, Korrawe and Preechakul, Konpat and Aksan, Emre and Beeler, Thabo and Suwajanakorn, Supasorn and Tang, Siyu},
  booktitle={Proceedings of the IEEE/CVF Conference on Computer Vision and Pattern Recognition},
  pages={1334--1345},
  year={2024}
}

@inproceedings{mu2025stablemotion,
  title={StableMotion: Training Motion Cleanup Models with Unpaired Corrupted Data},
  author={Mu, Yuxuan and Ling, Hung Yu and Shi, Yi and Ojeda, Ismael Baira and Xi, Pengcheng and Shu, Chang and Zinno, Fabio and Peng, Xue Bin},
  booktitle={Proceedings of the SIGGRAPH Asia 2025 Conference Papers},
  pages={1--12},
  year={2025}
}

@inproceedings{lugmayr2022repaint,
  title={Repaint: Inpainting using denoising diffusion probabilistic models},
  author={Lugmayr, Andreas and Danelljan, Martin and Romero, Andres and Yu, Fisher and Timofte, Radu and Van Gool, Luc},
  booktitle={Proceedings of the IEEE/CVF conference on computer vision and pattern recognition},
  pages={11461--11471},
  year={2022}
}

@InProceedings{humanML3D,
    author    = {Guo, Chuan and Zou, Shihao and Zuo, Xinxin and Wang, Sen and Ji, Wei and Li, Xingyu and Cheng, Li},
    title     = {Generating Diverse and Natural 3D Human Motions From Text},
    booktitle = {Proceedings of the IEEE/CVF Conference on Computer Vision and Pattern Recognition (CVPR)},
    month     = {6},
    year      = {2022},
    pages     = {5152-5161}
}

@article{rempe2026kimodo,
  title={Kimodo: Scaling Controllable Human Motion Generation},
  author={Rempe, Davis and Petrovich, Mathis and Yuan, Ye and Zhang, Haotian and Peng, Xue Bin and Jiang, Yifeng and Wang, Tingwu and Iqbal, Umar and Minor, David and de Ruyter, Michael and others},
  journal={arXiv preprint arXiv:2603.15546},
  year={2026}
}

\clearpage

\end{document}